\documentclass[sigconf]{acmart}
\AtBeginDocument{%
  }

\copyrightyear{2025}
\acmYear{2025}
\setcopyright{acmlicensed}\acmConference[ICPE Companion '25]{Companion of the 16th ACM/SPEC International Conference on Performance Engineering}{May 5--9, 2025}{Toronto, ON, Canada}
\acmBooktitle{Companion of the 16th ACM/SPEC International Conference on Performance Engineering (ICPE Companion '25), May 5--9, 2025, Toronto, ON, Canada}
\acmDOI{10.1145/3680256.3721973}
\acmISBN{979-8-4007-1130-5/2025/05}

\usepackage{graphicx}
\usepackage{subcaption}
\usepackage{makecell}
\usepackage{multirow}

\begin{document}

\title{A Dataset of Performance Measurements and Alerts from Mozilla (Data Artifact)}

\author{Mohamed Bilel Besbes}
\email{m\_besbes@live.concordia.ca}
\affiliation{%
  \institution{REALISE Lab @ Concordia University}
  \city{Montréal}
  \state{Québec}
  \country{Canada}
}

\author{Diego Elias Costa}
\email{diego.costa@concordia.ca}
\affiliation{%
  \institution{REALISE Lab @ Concordia University}
  \city{Montréal}
  \state{Québec}
  \country{Canada}
}

\author{Suhaib Mujahid}
\email{smujahid@mozilla.com}
\affiliation{%
  \institution{Mozilla}
  \city{Montréal}
  \state{Québec}
  \country{Canada}
}

\author{Gregory Mierzwinski}
\email{gmierzwinski@mozilla.com}
\affiliation{%
  \institution{Mozilla}
  \city{Potton}
  \state{Québec}
  \country{Canada}
}

\author{Marco Castelluccio}
\email{mcastelluccio@mozilla.com}
\affiliation{%
  \institution{Mozilla}
  \city{London}
  \country{United Kingdom}
}

\renewcommand{\shortauthors}{Mohamed Bilel Besbes, Diego Elias Costa, Suhaib Mujahid, Gregory Mierzwinski, \& Marco Castelluccio}
\begin{abstract}
  Performance regressions in software systems can lead to significant financial losses and degraded user satisfaction, making their early detection and mitigation critical. Despite the importance of practices that capture performance regressions early, there is a lack of publicly available datasets that comprehensively capture real-world performance measurements, expert-validated alerts, and associated metadata such as bugs and testing conditions. 

    To address this gap, we introduce a unique dataset to support various research studies in performance engineering, anomaly detection, and machine learning. 
    This dataset was collected from Mozilla Firefox's performance testing infrastructure and comprises 5,655 performance time series, 17,989 performance alerts, and detailed annotations of resulting bugs collected from May 2023 to May 2024. By publishing this dataset, we provide researchers with an invaluable resource for studying performance trends, developing novel change point detection methods, and advancing performance regression analysis across diverse platforms and testing environments.
    The dataset is available at~\url{https://doi.org/10.5281/zenodo.14642238}.
\end{abstract}



\maketitle

\section{Introduction}
\label{sec:introduction}

Software performance testing and monitoring have become crucial practices for ensuring the quality of modern software. 
Companies manage diverse platforms, and frequent updates to their products, and even small performance regressions can lead to significant financial losses and degraded user satisfaction. 
For instance, a study at Amazon revealed that a one-second delay in page load speed could result in an estimated \$1.6 billion loss in annual revenue\cite{amazon-1-6-b-loss}. 
Also, Amazon stated that every 100 milliseconds in added page load time cost 1\% in revenue in a study dating back to 2006\cite{amazon-100-ms-effect}. On the other hand, software performance improvements, even if slight, help businesses with increasing performance indicators such as customer retention. For example, in 2011, a study showcased that a page load time reduction by 2.2 seconds of the download page for Mozilla's famous browser, Firefox, resulted in an additional 10 million Firefox downloads in a single year\cite{mozilla-sitespect}. 

Maintaining efficient software performance has driven the adoption of robust performance engineering practices in the industry. 
Numerous companies spend significant time and resources to devise practices for detecting and mitigating performance regressions before they impact production~\cite{creatingvirtuouscyclepaper}.
There have been efforts in publishing industry datasets for performance measurement~\cite{strandberg2023westermotestperformancedata,hardt2024petshopdatasetfinding,davidicpe2022}. 
MongoDB's performance dataset that was published by Daly~\cite{davidicpe2022}, in the ICPE Data Challenge of 2022.
However, we believe that there is a lack of comprehensive datasets that capture these industrial practices that researchers can study.
We introduce a dataset that contains one year of performance testing and monitoring data from Mozilla.  
This dataset includes 5,655 time-series performance measurements, 17,989 performance alerts with expert validation, and their associated performance bugs.

The development of this dataset involved an important effort in collecting, cleaning, and annotating performance measurements from diverse software systems of Mozilla. 
To enhance its utility, the performance measurements data was cross-referenced with the alerts. 
Also, the industrial effort of conducting the performance testing and manually annotating it was very notable as performance alerts are carefully validated and linked to corresponding bugs. This would enable researchers to trace the root causes of regressions effectively. By providing high-quality, annotated data, this dataset serves as a valuable resource for analyzing performance trends, detecting performance anomalies, and improving platform-specific workflows.

\section{\textbf{Context and Background}}
\label{sec:perf_eng_at_mozilla}




Mozilla Firefox is a popular open-source internet browser, known for its emphasis on privacy and customization, used by millions of users worldwide~\cite{firefox}. 
As responsiveness is a critical quality of internet browsers, the development team employs robust performance testing and monitoring strategies to capture potential regressions before they reach their end users. 
As new code is pushed to the codebase, multiple performance tests are executed on periodic basis to assess the performance of the new code. 
For each code revision~\cite{mercurial_revision} under test, performance tests run on various \textit{platforms} (e.g., a specific operating system running on desktop hardware). 
The result of a single performance test is a \textit{performance measurement}, which can represent execution time, memory consumption, and other tested performance characteristics.   
A platform is a collection of software and hardware setup

Given the inherent variability of software performance~\cite{tamingsoftwarevaribility}, performance tests could be repeated multiple times to increase the robustness of the results, a common practice in software performance engineering~\cite{wwhatswrongwithmybenchmark}.
A \textit{performance time series} is a sequence of performance measurements of the same performance test on a single platform throughout multiple revisions, and is commonly identified as a \textit{signature} in Mozilla's terminology.



Mozilla's performance anomaly detection system is called Perfherder, and it is developed to identify performance anomalies that need further investigation. 
Mozilla's Perfherder periodically runs performance checks by analyzing one of the recent revisions by computing its T-test score from the two-sample student T-test\cite{student1908} to contrast the measurements of generally 12 preceding revisions minimum that had performance measurements with the current plus a maximum of 11 subsequent ones.
The resulting T-value is used as a confidence score showcasing the likelihood of the occurrence of a performance anomaly.
In case the T-test shows a significant change detected by comparing it to a fixed threshold, Perfherder proceeds with measuring the change magnitude between both measurement groups, and if it surpasses a certain threshold, an alert is triggered on Perfherder~\cite{perfherder}.
Perfherder is the full system used in Mozilla to handle the performance workflow. 

\begin{figure}[h!]  
    \centering

    \begin{subfigure}[b]{\columnwidth}
        \centering
        \includegraphics[width=\columnwidth]{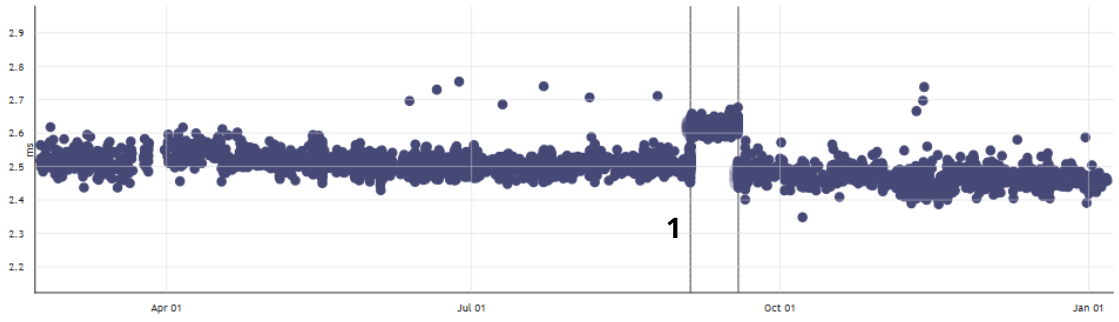} 
        \caption{Timeseries example}
        \label{fig:timeseries_example} 
    \end{subfigure}

    \begin{subfigure}[b]{\columnwidth}  
        \centering
        \includegraphics[width=\columnwidth]{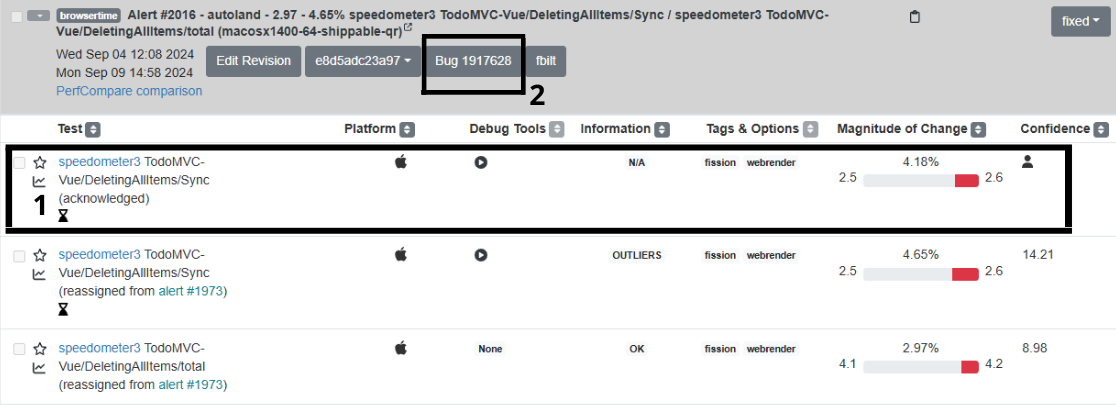} 
        \caption{Alert summary example}
        \label{fig:alert_summary_example} 
    \end{subfigure}
    
    \vspace{0.2cm}  

    \vspace{0.2cm}  

    \begin{subfigure}[b]{\columnwidth}
        \centering
        \includegraphics[width=\columnwidth]{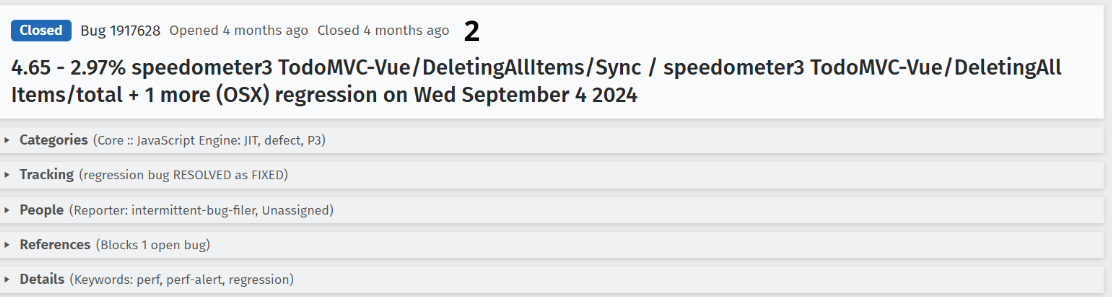} 
        \caption{Bug example}
        \label{fig:bug_example} 
    \end{subfigure}

    \caption{Illustrative example of the data as seen by Performance Sheriffs}
    \label{fig:timeseries_alert_summary_bug_example}
\end{figure}

Alerts related to the same software revision are grouped in an \textit{alert summary} (as presented in \textbf{1} in Figure \ref{fig:timeseries_alert_summary_bug_example}).
These alert summaries are then manually evaluated by a \textit{Performance Sheriff}, a member of Mozilla's performance team, who assesses whether the alert should be further investigated. 
Performance Sheriffs can also create new alerts manually, if they notice a performance anomaly that was missed by Perfherder. 
In case the investigated alert summary presents an actual regression, a bug is created and associated with the given alert summary as shown in \textbf{2} in the same figure.


\section{Dataset Collection and Processing}
\label{sec:dataset_collection_analysis}

\subsection{Artifacts \& dataset description}
\label{subsec:artif_dataset_desc}

\begin{figure}
    \centering
    \includegraphics[width=0.9\columnwidth]{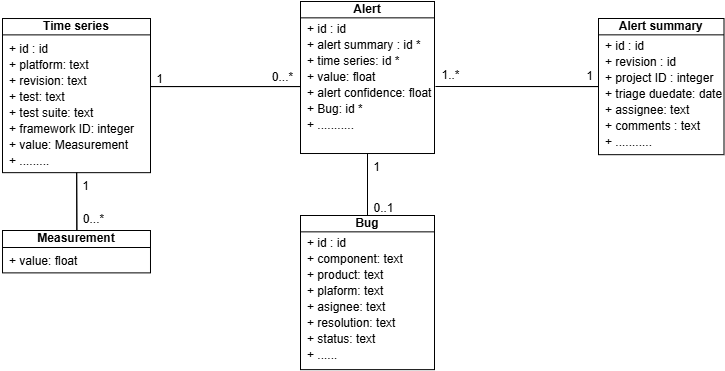}
    \caption{Structure of the relationships between the data entities}
    \label{fig:class_diagram}
\end{figure}

Our collected dataset contains four main entities: 
the performance time series, the alerts data, their associated alert summaries, and their associated bugs' data.
Figure \ref{fig:class_diagram} showcases the relationships between the entities. 

\noindent
\textbf{Performance time series:}
Performance time series contain series of performance measurements of a performance test and platform across multiple software revisions (x scale). 
The data contains all the measurement values, the measurement unit, the platform, its related test suite, and more. 
All time series collected contain at least one performance alert.


\noindent
\textbf{Alerts:}
An alert is a potential performance anomaly that was triggered by Mozilla Perfherder (automated) or by performance Sheriffs (manual). 
The alerts data characterizes the performance alert and contains the related alert summary (grouping), the alert's status, the time series ID, the result of the t-test, whether the alert was created manually or not, the noise profile of that alert if there is any, and other attributes.

\noindent
\textbf{Alert Summaries:} 
An alert summary is a grouping of alerts on the same software revision. A revision could have multiple associated alert summaries.
It is important to note that, alert summaries are the ones usually validated by Performance Sheriffs. 
The alert summary data contains details about its related software revision, the triage due date, the assignee among the performance Sheriffing team that will take a look at the alert summary, its associated bug, and other attributes.

\noindent
\textbf{Performance Bugs data:}
If a performance alert summary is validated by Performance Sheriffs, usually a performance bug is created to prompt the development team to action. 
The performance bug entity holds information about bug creation metadata (e.g., time of the creation, author), bug severity, bug status, creator, assignee, its associated product and component, comments from Sheriffs, its status of replication, and other attributes.


\noindent

\subsection{Data collection process}
\label{subsec:data-collection-process}



We collect the entire dataset using the Mozilla Perfherder API~\cite{mozilla-api}.
This particular API retains the historical data of alerts and performance measurements for one year, hence, we collect the data from May 2023 to May 2024.
We proceed to collect the data using the following steps: 

\begin{enumerate}
    \item We started by collecting performance alerts present in the Mozilla API, relevant to all Mozilla systems. 
    
    \item Using the collected performance alerts, we identified the unique time series related to these alerts and we extracted their features.
    This means that all of our performance time series are associated with at least a single alert, and we have also not included time series that have not exhibited at least one single performance alert. 

    \item We cleaned the dataset by removing empty time series for example. We also cross-referenced the alerts features with the time series measurements.
    
    \item Similar to the time series, we identified the unique bugs associated with the collected alerts and we extracted their features.
    
\end{enumerate}


\subsection{Data Labeling}
\label{subsec:status_categorization}



Alert summaries group multiple alerts from the same software revision and are validated by Performance Sheriffs. These alerts could be created by Perfherder or by Performance Sheriffs.
Upon validation, Performance Sheriffs update the alert summary status, which indicates whether an alert summary was deemed to be false (e.g., environmental change unrelated to the software) or needs further action from their team. 
We wanted to facilitate the use of this dataset by the research community, hence, we provide a custom labeling system, as follows:


\begin{itemize}
    \item \textbf{True Alert summaries:} True alert summaries represent validated performance anomalies or regressions. They include alerts from summaries labeled as \textit{reassigned}, \textit{improvement}, \textit{fixed}, \textit{backedout}, \textit{downstream}, or \textit{wontfix}.

    \item \textbf{False Alert summaries:} False alert summaries do not represent real performance issues. They have the \textit{invalid} status. They result from noise or irrelevant factors, for example.

    \item \textbf{Uncertain Alert summaries:} These are alert summaries whose validity remains undetermined. They are labeled as \textit{investigating} or \textit{untriaged}, requiring further review to classify them as true or false.
\end{itemize}

\subsection{Dataset Characteristics}

Table~\ref{tab:dataset_summary_stats} showcases some of the statistics of our dataset, including performance time series, their associated alerts and performance bugs. 
Our dataset contains a total of 5,655 performance time series, covering performance tests of 186 different test suites across 5 different software platforms. 
On average, each performance time series contains 2,124 measurements and tests, and 1,867 software revisions.  
In total, the dataset contains $\approx$12 million measurements, and only 0.35\% of performance measurements are associated with a performance alert.

Analyzing the performance alerts, we report a total of 17,989 alerts, 8,788 of which correspond to alerts from Speedometer3/TP6 test suites, the two of the most important Mozilla performance test suites. 
When grouped, the alerts correspond to 3,912 alert summaries.
Most of the alert summaries come from two repositories, Autoland \cite{autoland} and Mozilla Beta \cite{mozilla-beta}, respectively, representing 84.4\% and 12.06\% of the total alert summaries. 
Autoland is the first testing stage, followed by Mozilla Beta, which justifies the high occurrence of performance alerts in Autoland versus Mozilla Beta. 
The breakdown of the repository distribution is detailed in Table \ref{tab:dataset_summary_stats}.

The total of 3,912 alert summaries are categorized as follows according to the logic stated in \ref{subsec:status_categorization}. True alerts represent 56.31\% of the alerts, Uncertain alerts represent 31.21\%, and False alerts represent 12.47\%. Table \ref{tab:dataset_summary_stats} contains the alert summaries count per status for every single project.

\begin{table*}
\caption{Dataset statistics by repository.}
\label{tab:dataset_summary_stats}
\centering
\begin{tabular}{lrr|rrr|rr}
\toprule
& 
&
&
\multicolumn{3}{c|}{\textbf{Custom Alert Labels}}
& 
& 
\\

\textbf{Repository} & 
\textbf{Time Series} &
\textbf{Alerts} & 
\textbf{True} & 
\textbf{False} & 
\textbf{Uncertain} & 
 \textbf{Alert Summaries} & 
\textbf{Bugs} \\

\midrule

Autoland & 3833 & 13593 & 1779 & 431 & 1109 & 3319 & 438 \\ 
Mozilla Beta  & 1477 & 3597 & 317 & 45 & 110 & 472 & 25 \\ 
Firefox Android & 342 & 796 & 105 & 12 & 1 & 118 & 25 \\ 
Mozilla Central & 2 & 2 & 2 & 0 & 0 & 2 & 2 \\ 
Mozilla Release & 1 & 1 & 0 & 0 & 1 & 1 & 1 \\ 
\midrule
\textbf{Total} & \textbf{5655} & \textbf{17989} & \textbf{2203} & \textbf{488} & \textbf{1221} & \textbf{3912} & \textbf{482} \\ 

\bottomrule

\end{tabular}
\end{table*}

\begin{table}
\caption{Breakdown of the Alerts/Alert Summaries by Platform}
\label{tab:platform_summary_stats}
\centering
\begin{tabular}{lrr}
\toprule
\textbf{Platform} & \textbf{Alerts} & \textbf{Alert Summaries} \\
\midrule
Windows & 6241 & 1754 \\
macOS & 5298 & 1615 \\
Linux & 4210 & 1302 \\
Android & 2217 & 330 \\
Other & 23 & 23 \\
\midrule
\textbf{Total}    & \textbf{17989} & \textbf{3912}           \\ 
\bottomrule
\end{tabular}
\end{table}

\begin{table}[]
    \caption{Characteristics of the 482 performance bugs in the dataset}
    \label{tab:bug_characteristics}
    \centering
    \begin{tabular}{llr}
    \toprule
        \textbf{Characteristic} & \textbf{Level} & \textbf{\# of Bugs} \\
    \midrule
        \multirow{4}{*}{Bug Severity} 
& S2 & 26 \\
& S3 & 125 \\
& S4 & 33 \\
& Unknown & 298 \\
         \midrule
        \multirow{5}{*}{Bug Status}

& NEW & 35 \\
& ASSIGNED & 3 \\
& REOPENED & 1 \\
& RESOLVED & 427 \\
& VERIFIED & 16 \\
\bottomrule
    \end{tabular}
\end{table}


To test the performance of a software revision, performance tests are executed across different software platforms. 
We report the alerts per tested platforms in Table \ref{tab:platform_summary_stats}. We also count the presence of a given platform at least once in alert summaries.
There is a large presence of alerts from tests that run on Windows (6,241 alerts), macOS (5,298 alerts), Linux (4,210 alerts), and Android (2,217 alerts).  


In case a performance regression is identified, it gets associated to a bug to be fixed. It is worth noting that out of the 3,912 existing alert summaries, only 633 have one associated bug out of the 482 unique bugs. Table \ref{tab:bug_characteristics} contains the breakdown of bugs by severity and status. The severity of the bug decrease in magnitude as the numerical value increases.


\noindent
\textbf{Structure of the dataset artifact.}
The dataset artifact~\cite{replicationpackage} is organized as follows:

\begin{itemize}
    \item \textbf{Scripts:} A folder that contains the scripts used to collect, clean and label the data. The scripts can also be used to re-collect and update the dataset to include newer performance measurements and alerts from Mozilla systems.  

    \item \textbf{Data:} A folder containing the performance time series data, alerts and alerts summaries, and bugs. Performance time series are further organized into their respective repositories, and we store a single CSV per performance time series. Alerts and bugs, on the other hand, are stored on one CSV file each.  

\end{itemize}

\section{Areas of data potential usage}
\label{sec:potential_data_usage}

Given that the dataset includes performance time series, validated performance alerts, and their associated bugs, we envision that it can be used to conduct research on different areas in performance engineering. 

\noindent
\textbf{Performance characterization}
The performance time series can be used to best understand the performance profile of industrial software. 
Researchers can use this dataset to characterize the performance evolution of Mozilla systems, identify performance trends, characterize performance measurement variation, and correlate measured performance across different testing platforms.


\noindent
\textbf{Performance Regression Prediction}
The dataset includes performance measurements, expert-validated performance changes (alerts), and our custom label that makes it ideal for predicting performance regression. 
This dataset can be used to test different approaches, such as change point detection methods~\cite{CPD}, time-series forecasting~\cite{10.1145/3629526.3645049} or regression models~\cite{10.5120/3840-5341}.
The dataset also contains other metadata attributes (such as the noise profile), which can further help classify performance measurements. 

\textbf{Characterization of Performance Bugs}
We envision researchers using the published dataset as a starting point to characterize performance bugs. 
The dataset contains metadata that can be used to assess how long bugs take to be created and fixed, how many professionals are directly involved in the related discussion, and how much debate goes into fixing a performance bug.

\textbf{Performance regressions root cause analysis}
Given Mozilla develops open-source software, it is also possible to extend this work to extract the code-related features. 
The meta-data included in the dataset targets both the repository, test suite, and the revision ID, that can be used to mine the exact commit and 
to analyze what are the code modifications that caused regressions. This is especially promising because the dataset covers multiple repositories/projects.

\section{Limitations}
\label{sec:limitations_future_works}

\textbf{Bias in the data collection:} we have collected performance measurements associated with at least a single performance alert. 
Performance time series that do not yield any alert is not present in the dataset. So, the dataset might not fully represent the entire spectrum of performance measurements, particularly normal or less 'anomalous' measurements that do not trigger alerts. Researchers can, however, use our data collection scripts to expand the dataset to include times series that do not exhibit any anomalous behavior. 

\textbf{Potential threats in data collection:} Data alerts and associated performance measurements were collected on different dates. This temporal separation could lead to minor inconsistencies in the dataset. 
We mitigated this threat by cross-verifying the alerts and measurements alignment. 

\textbf{Generalizability:} While the dataset focuses on Mozilla-tested products, including browsers and other software, it may not generalize seamlessly to products outside the Mozilla ecosystem. 
However, we believe that many practices and processes used by Mozilla’s performance team (such as time series monitoring, alerting mechanisms, and regression identification strategies) can offer transferable insights to other use cases. 



\section{Conclusion and Future work}

We present a novel dataset of performance time series, alerts and bugs, tailored to support research on software performance engineering. 
This dataset is collected from real-world software performance monitoring systems in Mozilla industrial settings. 
By providing this resource, we hope to enable further studies in a variety of performance engineering topics.


In the near future, we aim to extend the dataset by including the latest performance measurements and potentially include code-related features. 
We encourage the community to build on this artifact in order to push the boundaries of performance engineering research and explore new methodologies for ensuring software reliability and efficiency.



\end{document}